\documentstyle[aps]{revtex}
\newcommand{\be}{\begin{equation}}
\newcommand{\ee}{\end{equation}}
\newcommand{\bea}{\begin{eqnarray}}
\newcommand{\eea}{\end{eqnarray}}

\font\mybb=msbm10 at 10pt
\def\bb#1{\hbox{\mybb#1}}

\def\bE {\bb{E}}

\def\e  {\epsilon}

\def\tr{\rm tr}

\begin{document}

\twocolumn[\hsize\textwidth\columnwidth\hsize\csname
@twocolumnfalse\endcsname
\rightline{DAMTP-1999-67}
\rightline{hep-th/9905156}
\title{Supersymmetry and Generalized Calibrations}
\author{J. Gutowski, G. Papadopoulos and P.K.
Townsend}
\address{DAMTP, Univ. of Cambridge, Silver St.,\\
Cambridge CB3 9EW,  UK\\}

\maketitle
\begin{abstract}
A static minimal energy configuration of a 
super p-brane in a supersymmetric
(n+1)-dimensional spacetime is shown to be a `generalized
calibrated' submanifold. Calibrations in 
$\bE^{(1,n)}$ and $AdS_{n+1}$ are special
cases. We present several M-brane examples.

\end{abstract}
\vskip2pc]
\section{{\bf INTRODUCTION}}
A super p-brane \cite{pol,bst,ach} is a 
charged p-dimensional object moving in
superspace, with a charge equal to its 
tension, which we set to unity. If the
superspace is a supersymmetric one then 
the (n+1)-dimensional spacetime will
admit a timelike Killing vector field 
$k$ and the motion will conserve a
corresponding energy $E$. The fact that 
the p-brane is charged means that it
couples to a superspace (p+1)-form 
potential $A$. There are therefore  two
contributions to the energy density 
of a static brane, one proportional to the
p-volume density, the other an 
`electrostatic' energy density. The p-volume
density is $\sqrt{\det m}$ where 
$m$ is the induced worldspace p-metric. The
electrostatic energy density is 
the worldspace dual of the p-form induced by
$i_kA$; let us call this $\Phi$. 
Thus, as we shall confirm below, stable static
bosonic solutions to the worldvolume 
field equations (the `branewave' equations)
minimize a potential energy functional of the form
\be\label{enfun}
E = \int d^p\sigma \left[ \nu \sqrt{\det m} + \Phi
\right]\, ,
\ee
where $\sigma^i$ ($i=1,\dots,p$) are the
 worldspace coordinates, and $\nu$ is a
redshift factor; for static spacetimes 
$\nu = \sqrt{-k^2}$. We shall call these
static configurations `minimal worldspaces'. 
For a vacuum superspace
background the spacetime is (n+1)-dimensional 
Minkowski space, $\bE^{(1,n)}$. 
In this case $\Phi=0$ because the (p+1)-form 
$A$ is purely fermionic. In
addition, $\nu=1$, so in this case a 
minimal worldspace is a minimal surface in
$\bE^n$ in the usual sense. We see 
that super p-branes provide a natural
framework for an extension of the 
notion of a minimal surface in $\bE^n$ to
minimal worldspaces in more general Riemannian spaces.  

Supersymmetry did not play a major role 
in the above discussion and it might be
deemed an unnecessary complication. 
However, a large class of minimal surfaces 
in $\bE^n$ are calibrated surfaces 
for which the calibrating p-form, 
or `calibration' \cite{hl}, admits a 
spinorial construction \cite{dh}. Let $x^m=(x^0,x^I)$
be the Minkowski spacetime coordinates, 
with $x^I$ ($I=1,\dots,n$) being
cartesian coordinates on $\bE^n$. 
Then, for $p=1,2$ mod $4$, the p-form
calibration takes the form
\be\label{pform}
\varphi = dx^{I_1} \wedge \dots \wedge dx^{I_p} \epsilon^T
\Gamma_{0I_1\dots I_p}\epsilon 
\ee
where $\epsilon$ is a constant 
real spinor\footnote{The 
requirement that $\epsilon$ be real
restricts the spacetime dimension, and 
hence the space dimension $n$; the extent
to which the constraints on $p$ and 
$n$ may be relaxed will be discussed
below.} normalized so that
\be
\epsilon^T \epsilon =1\, ,
\ee
and $\Gamma_{m_1\dots m_k}$ are 
antisymmetrized products of the 
Minkowski spacetime Dirac matrices. Many calibrations of this form 
have been shown \cite{calib} to calibrate
supersymmetric intersecting brane 
configurations in M-theory  (for which $\epsilon$ is a real
32-component spinor of the 11-dimensional Lorentz group). 
Similar calibrations were earlier discussed in the context of
Calabi-Yau compactifications of string theory \cite{ooguri}.
The essential point is
that, given a  tangent p-plane $\xi$, we can write
$\varphi|_\xi$ as
\be\label{pformb}
\varphi|_\xi = \sqrt{\det m}\, \epsilon^T \Gamma_\xi \epsilon
\ee
where $\Gamma_\xi$ is the matrix 
\be\label{gamma}
\Gamma =  {1\over p! \sqrt{\det m}}\,\varepsilon^{i_1\dots i_p}
\partial_{i_1}x^{I_1}\cdots
\partial_{i_p}x^{I_p} \Gamma_{0I_1\dots I_p}\, ,
\ee
evaluated at the point to which $\xi$ is tangent. Given the above
restriction on the values of $p$, we have 
\be\label{gamcon}
\Gamma^2=1 \, .
\ee
It follows that $\varphi|_\xi \le vol_\xi$ 
for all $\xi$. Since $\varphi$ is
also closed this means that it is a 
calibration. The `contact set' of the
calibration is the set of p-planes for 
which this inequality is saturated.
By the theorem of Harvey and Lawson, 
these are the tangent planes of a minimal
surface \cite{hl}.

We see from (\ref{pformb}) that an 
alternative characterization of the contact
set is that set of p-planes $\xi$ for which
\be\label{contact}
\Gamma_\xi \epsilon =\epsilon \, .
\ee
Because of (\ref{gamcon}) this equation 
is automatically satisfied for any
{\sl given} tangent p-plane $\xi$ and, 
since $\tr \Gamma=0$, the solutions
$\epsilon$ belong to a `1/2 supersymmetric' 
subspace of spinor space. However,
this solution space generally varies as 
$\xi$ varies over the contact set so the
solution space of the set is generally 
smaller. The connection of calibrations
to supersymmetry observed in \cite{calib} 
is based on the fact that the matrix
$\Gamma$ is closely related to a matrix 
with similar properties appearing in the
$\kappa$-symmetry transformations of the 
worldvolume fields of the super p-brane
in its Lagrangian formulation. In fact, for 
time-independent worldvolume fields
in the gauge
$x^0=t$ the latter matrix reduces to 
the one of (\ref{gamma}), but in this
context the condition (\ref{contact}) is 
just the condition for preservation of
some fraction of supersymmetry; the 
total number of preserved supersymmetries is
the dimension of the solution space for 
$\epsilon$. Note that although the
specific form of the calibrating p-form 
given above required restrictions on
both $p$ and $n$, the results can be 
extended without difficulty to all those
pairs $(p,n)$ for which there exists a 
super p-brane action. The possibilities
were classified in \cite{ach} under 
the assumption that the only bosonic
worldvolume fields are scalars, and 
these are the cases that we have in mind
here. The 11-dimensional supermembrane 
\cite{bst} is a simple example with $(p,n)=(2,10)$. 
Of course, many of
the results to be obtained will 
apply to other branes such as D-branes and the
M5-brane if one sets to zero any 
non-scalar worldvolume fields. In fact, most
discussions to date of the relevance of
calibrations to branes have been in the
context of the M5-brane. 

Most previous discussions have also concentrated on the case of
branes in a flat Minkowski background. In considering the extension 
to non-flat backgrounds it is convenient to distinguish between those
for which the energy functional of a p-brane is still equal to the integral of
the p-volume and those for which the more general formula (\ref{enfun}) is
required. In the former case, a static p-brane is again a minimal surface and
supersymmetric minimal p-surfaces are calibrated by p-form calibrations in
much the same way as in the above discussion. The examples of
\cite{ooguri} are in this category. As we shall show here,
the recently discussed \cite{bergtow} `lump'
solitons on the M2-brane in an M-monopole background provide
further examples of K\"ahler calibrations of this type, and 
there are a number of
related K\"ahler calibrations that we discuss. If, on the other hand, 
a static brane of minimal
energy is {\sl not} a minimal surface then it cannot be a calibrated
surface in the sense of \cite{hl}, but 
it may still be a calibrated surface in a
more general sense because it may be calibrated by a form that
is {\sl not} closed; examples are provided by the `adS' calibrations
\cite{ads} which calibrate p-surfaces of minimum p-brane energy in anti-de
Sitter backgrounds. The adS examples suggest a definition of a `generalized
calibration', which we present here; we then show, by a straightforward
extension of an argument in \cite{ads}, that p-surfaces calibrated by such
generalized calibrations are `minimal worldspaces'. We discuss a number of
examples provided by M-branes in 1/2-supersymmetric M-theory backgrounds. 

Another purpose of this article is to show how the connection between
calibrations and super p-branes follows directly from the
supersymmetry algebra of Noether charges \cite{AGIT,sato}. 
In particular, we shall see that all the properties of generalized calibrations
arise naturally in this context. For this reason we shall begin with this
analysis. Since the supersymmetry charges are integrals of functions on phase
space, it is natural to start from the Hamiltonian formulation, the details of
which can be found in \cite{berg,ait,drev}. Having defined the notion
of a generalized calibration and proved the associated minimal worldspace
property of calibrated surfaces, we then move on to the explicit examples.

\section{{\bf THE SUPERSYMMETRY ALGEBRA AND THE CALIBRATION BOUND}}
We begin by making precise the constructions 
alluded to in the opening paragraph.
Let $Z^M=(x^m,\theta^\mu)$ be the 
superspace coordinates. The geometry of
superspace will be assumed to be 
determined by the Lorentz frame 1-forms
$E^A=(E^a,E^\alpha)$ and the (p+1)-form 
potential $A$; we will assume
throughout that the only non-vanishing 
component fields are bosonic.
The superspace (p+1)-form $A$ appears in 
the Lagrangian form of the  action via
its pullback to the worldvolume and the 
integrand can be written in the form
$\dot Z^M {\cal A}_M$ where the dot 
denotes differentiation with respect to an
arbitrary worldvolume time $t$; the 
coefficients ${\cal A}_M$ are the components
of the one-form ${\cal A}$ obtained by 
contraction of $A$ with the $p$ tangent
vectors spanning a tangent p-plane $\xi$ to 
the p-dimensional worldspace, i.e.
${\cal A}=A|_\xi$. Now let $x^m=(x^0,x^I)$, 
where $k=\partial/\partial x^0$ is
the timelike Killing vector field, and 
fix the time reparameterization
invariance by choosing $x^0=t$. Then
\be
\dot Z^M {\cal A}_M = \Phi + \dot x^I {\cal A}_I 
+ \dot\theta^\mu {\cal A}_\mu\, 
\ee
where $\Phi$ is a superfield with the 
electrostatic potential density as its
lowest component. As is customary, we use the same 
symbol to denote a superfield and its lowest
component. Note that $\Phi$ is the 
worldspace dual of the pullback of $i_kA$. The
worldspace metric is $m_{ij} 
= E_i^a E_j^b\eta_{ab}$ where $E_i{}^a= \partial_i
Z^M E_M{}^a$ and $\eta$ is the 
Minkowski metric. Omitting fermions it reduces,
in the $x^0=t$ gauge, to
\be
m_{ij} = \partial_ix^I \partial_j x^J g_{IJ}
\ee
where $g_{IJ}$ is the metric on a 
spatial section of spacetime, 
i.e. the p-metric on worldspace is 
induced from the n-metric on space.

The super p-brane Lagrangian density can be written as 
\be\label{full}
{\cal L} = \dot Z^MP_M - s^i E_i{}^a\tilde p_a -
{1\over2}\lambda \left(\tilde p^2 + \det m\right)
\ee
where $s^i$ and $\lambda$ are Lagrange multiplers, and
\be
\tilde p_a = E_a{}^M\left( P_M + {\cal A}_M\right)\, .
\ee
This is not its fully canonical form because
$\kappa$-symmetry implies a constraint 
for which there is as yet no
corresponding Lagrangian multiplier. However,
 this constraint affects only the
fermionic variables. If we set all
 worldvolume fermions to zero and use the
identity $\partial_i x^m {\cal A}_m\equiv 0$ then 
\be
{\cal L} = \dot x^mp_m - s^i \partial_i x^m p_m -
{1\over2}\lambda \left(\tilde p^2 + \det m\right)
\ee
where, now, $\tilde p^2 
=g^{mn}\tilde p_m \tilde p_m$, with $\tilde p_m = p_m +
{\cal A}_m$ and $g^{mn}$ the inverse of the 
spacetime metric $g_{mn}= E_m{}^a
E_n{}^b\eta_{ab}$. Now set 
\be
p_m = (-{\cal H},P_I)\, ,
\ee
and impose the $x^0=t$ gauge, to get
\be
{\cal L} = \dot x^I p_I - {\cal H} - s^i \partial_i x^I p_I
\ee
where the Hamiltonian density ${\cal H}$ is 
found by solving the Hamiltonian
constraint $\tilde p^2 + \det m=0$. This yields
\be
{\cal H}= \Phi + v^I\tilde p_I 
+ \nu \left[ g_{(n)}^{IJ}\tilde
p_I \tilde p_J + \det m\right]^{1/2}
\ee
where
\be
v^I = {g^{tI}\over (g^{tt})} \, , 
\qquad \nu = 1/\sqrt{-g^{tt}}\, ,
\ee
and $g_{(n)}^{IJ}$ is the inverse of 
the spatial n-metric $g_{IJ}$. For
simplicity we now restrict to static 
spacetimes, for which $v^I$ vanishes. In
this case $\nu =\sqrt{-k^2}$. 
For static worldvolume configurations 
we may set $\tilde p_I=0$  and the
Hamiltonian density reduces to 
\be
{\cal H}= \Phi + \nu \sqrt{\det m}\, .
\ee
Because $k=\partial_t$ is Killing, 
$m_{ij}$ and $\Phi$ are time-independent.
The integral of ${\cal H}$ over the 
worldspace yields the potential energy
functional $E$ of (\ref{enfun}). 

Supersymmetries of the background
 are associated with Killing
spinor fields. In many simple cases 
these take the form of a function times a
{\sl constant} spinor $\epsilon$ 
subject to a constraint of the form 
$\Gamma_*\epsilon =\epsilon$ where 
the constant matrix $\Gamma_*$ is
characteristic of the background. In 
particular, all 1/2-supersymmetric
backgrounds have Killing spinors of 
this form with $\Gamma_*$ a 
traceless matrix such that $\Gamma_*^2=1$. 
We shall assume here that the
background superspace is of this type. 
In addition, we continue to make the
inessential, but simplifying, assumptions 
that $p=1,2$ mod $4$, and that spinors are
real. In these cases
we may take the charge conjugation matrix to 
equal $\Gamma^{\underline 0}$, where
the underlining indicates a constant Lorentz 
frame matrix. The matrix
$\Gamma$ of (\ref{gamma}) can be generalized 
to non-flat spacetimes by
taking
\be
\sqrt{\det m}\,\Gamma =  \Gamma_{\underline 0} \gamma
\ee
where 
\be
\gamma = {1\over p!}\,\varepsilon^{i_1\dots i_p}
\partial_{i_1}x^{I_1}\cdots
\partial_{i_p}x^{I_p} \Gamma_{I_1\dots I_p}\, .
\ee
Note that, given the restriction on the values of $p$, we have
\be
\gamma^2 =-(-1)^p \det m\, .
\ee
Given our assumptions about the background 
the supersymmetry charges will take
the form of spinor functionals $Q_\alpha$ 
subject to the constraint
$Q\Gamma_* =Q$.
We will proceed by taking this projection
to be implicit,
and checking consistency at the end. 
With this understood we can write the
supersymmetry anticommutator as 
\be\label{alg}
\{Q,Q\} = \Gamma^{\underline 0}
\int d^p\sigma\,\nu\left[ \Gamma^a \tilde p_a + 
 \gamma\right]\, .
\ee

This result has not been directly 
established in full generality but it agrees
with the flat space case \cite{AGIT} and 
those non-flat cases that have been
analyzed \cite{sato}. That it must be 
correct follows indirectly from
considerations of $\kappa$-symmetry,
 as will now be explained. 
The $\kappa$-symmetry transformations 
were given in the required form in
\cite{ait}. In particular, the transformations 
of the fields $Z^M$ are such that 
$\delta_\kappa Z^M E_M{}^a =0$ and 
\be
\delta_\kappa Z^M E_M{}^\alpha = 
\left[\Gamma^a\tilde p_a - (-1)^p
\gamma\right]\kappa\, .
\ee
Since $E_m{}^\alpha$ vanishes for a bosonic 
background, and $E_\mu{}^\alpha
=\delta_\mu {}^\alpha$ at $\theta^\mu=0$, we
 may write the $\kappa$-symmetry
transformation as
\be
\delta_\kappa \theta = \left[\Gamma^a\tilde p_a -
 (-1)^p\gamma\right]\kappa\, .
\ee
This gauge transformation allows half the
 worldvolume fermions to be gauged
away. The supersymmetry transformations 
of the remaining half take the form
\be
\delta_\epsilon \theta = \epsilon 
+ \delta_{\kappa(\epsilon)}\theta
\ee
where $\kappa(\epsilon)$ is now a specific 
matrix function of worldvolume fields
acting on the constant spinor $\epsilon$. 
Its precise form depends on
the $\kappa$-symmetry gauge choice and it has the effect that
$\delta_\epsilon\theta = M\epsilon$ for 
some matrix $M$ \cite{itoh}. The matrix
$M$ depends on the worldvolume fields 
and hence, for static fields, on the
worldspace coordinates. At any given 
point on worldspace it is a matrix
for which half the eigenvalues vanish. 
This is due to the fact that,
locally, the brane realizes half the 
supersymmetry non-linearly. 
The spinors $\epsilon$ corresponding 
to the linearly-realized 
supersymmetries are the eigenspinors
 of $M$ with zero eigenvalue. These are the
supersymmetries preserved, locally, 
by the brane. The supersymmetries
that are globally preserved correspond to 
those spinors $\epsilon$ that are
eigenspinors of $M$ with vanishing 
eigenvalue for every point on worldspace.
Thus a worldvolume configuration is 
supersymmetric if the equation
$M(\sigma)\epsilon =0$ has solutions 
for constant $\epsilon$. This is 
equivalent to the equation 
$\delta_\epsilon \theta=0$. 

We now note the identity
\be
\left[\Gamma^a \tilde p_a +  \gamma\right]\delta_\kappa \theta
= \left[(\tilde p^2 + \det m)  - 2(\partial_i x^I p_I)
\gamma^i\gamma\right]\kappa
\ee
where $\gamma^i=m^{ij}\partial_jx^I\Gamma_I$.
The right hand side vanishes as a result of the Hamiltonian and
worldspace diffeomorphism constraints. 
Using these constraints we deduce that 
$\delta_\epsilon \theta=0$ implies
\be
\left[\Gamma^a \tilde p_a +  \gamma\right]\epsilon =0\, ,
\ee
which is an alternative form of the 
condition for preservation of supersymmetry
that obviates the need to compute the 
matrix $M$. This is the Hamiltonian form
of the supersymmetry condition derived in 
\cite{bbs,BKOP} from the Lagrangian
formulation. Observe that the matrix 
multiplying $\epsilon$ is precisely the
matrix appearing in the $\{Q,Q\}$ anticommutator, 
as required for consistency
because it is manifest from (\ref{alg}) that
 eigenspinors of this matrix with
vanishing eigenvalue correspond to 
supersymmetries that are preserved by the
p-brane configuration. 

We now turn to the consequences of the 
algebra (\ref{alg}) for static
configurations with $\tilde p_I=0$.  Given 
the normalized spinor $\epsilon$ of
the earlier discussion we deduce from 
the positivity of $\{Q,Q\}$ that
\be\label{bound}
-\tilde p_0 - \nu\, \epsilon^T 
\Gamma_{\underline 0}\gamma \epsilon \ge 0\, .
\ee
The first term on the left hand side is 
just $\nu \sqrt{\det m}$, by the
Hamiltonian constraint. For the second term we have
\be
\nu \epsilon^T\Gamma_{\underline 0} \gamma\epsilon  = \varphi|_\xi
\ee
where $\varphi|_\xi$ is the evaluation on a 
tangent p-plane $\xi$ of a p-form
$\varphi$ that is formally identical to 
(\ref{pform}) but now with $\Gamma_0 =
\nu \Gamma_{\underline 0}$ and Dirac matrices $\Gamma_I$ 
satisfying $\{\Gamma_I,\Gamma_J\}=2g_{IJ}$. 
Thus, (\ref{bound}) is equivalent to
\be\label{bo}
\varphi|_\xi \le  \nu\, vol_\xi \, .
\ee
 
Setting $\tilde p_I=0$, it follows from (\ref{alg}) that
\be
2(Q\epsilon)^2 = \int d^p\sigma \left[ {\cal H}- \left( 
i_k A + \varphi\right)|_\xi \right]\, .
\ee
The terms in the integrand other than ${\cal H}$ constitute a central 
extension of the supertranslation algebra and it follows from general
considerations that such extensions are topological. In other words, 
the p-form $i_k A +\varphi$ must be closed. Equivalently
\be\label{deriv}
d\varphi = {\cal F} \, ,
\ee
where ${\cal F}=-d i_kA$.
A p-form $\varphi$ that satisfies both the 
inequality (\ref{bo}) and the condition
(\ref{deriv}) will be called a `generalized calibration'. This definition is
modeled on the definition of an `adS' calibration in \cite{ads}, and adS
calibrations are examples of `generalized' calibrations; we will later provide
some further examples that are not adS calibrations. 

It will now be shown that the calibrated manifolds of a generalization
calibration are minimal worldspaces; the argument is a straightforward
extension of the one given for adS calibrations in \cite{ads}. Let
$U$ be an open subset of a calibrated manifold of a generalized calibration
$\varphi$. Then
\be
\int_U d^p\sigma\, \nu\sqrt{\det m} = \int_U\varphi
\ee
Now let $V$ be another open p-dimensional 
set in the same homology class as $U$ with $\partial
V=\partial U$, and let $D$ be a (p+1)-surface 
with $\partial D=U-V$. Then
\be
\int_U\varphi = \int_V \varphi + \int_D d\varphi\, .
\ee
Using (\ref{deriv}) we have
\be
\int_D d\varphi = -\int_D d (i_kA) = \int_V d^p\sigma\, \Phi  
-\int_U d^p\sigma\, \Phi
\ee
and hence
\bea
\int_U d^p\sigma \left[\nu \sqrt{\det m} + \Phi\right] &=& 
\int_V \varphi + \int_V d^p\sigma\, \Phi \nonumber
\\
\le \int_V d^p\sigma \left[\nu \sqrt{\det m} + \Phi\right]
\eea
where the inequality follows from (\ref{bo}). We conclude that 
\be\label{boun}
E(U)\le E(V)
\ee
where $E$ is the energy, and hence that $U$, initially defined as a subset of
the contact set of a generalized calibration, is also a subset of a minimal
worldspace. Since $U$ is arbitrary we conclude that the entire 
calibrated manifold of a generalized calibration is a minimal 
worldspace, as claimed.  We remark that if $di_kA=0$, then the 
above bound of the generalized calibration reduces to the bound for 
calibrated manifolds saturated by minimal surfaces.

Before proceeding to the examples we should discuss one potential difficulty
with the application of the above ideas. In many cases of interest, static
branes are invariant not only under time translations but also 
under some number, 
$\ell$ say, of space translations. In fact, in these cases the invariance group
is usually the Poincar\'e group of isometries of $\bE^{(1,\ell)}$, in which
case the brane configuration has an interpretation as an $\ell$-brane soliton
on the p-dimensional worldspace, i.e. an $\ell$-brane `worldvolume soliton'.
For $\ell>0$, the energy density is independent of some directions and the
total energy will be proportional to the volume of $\bE^\ell$, which is
infinite. This problem can be resolved by periodic identification of the
$\ell$ coordinates. Alternatively, we may minimise the energy
per unit $\ell$-volume, which we also denote by $E$. This is
\be
E=\int d^{p-\ell}\sigma 
\left [\nu_\ell \sqrt{{\rm det}\, m}+\Phi_\ell\right]\ ,
\ee
where (for static spacetimes) $\nu_\ell=\sqrt{ {\rm det}\, m_\ell}$, with 
$m_\ell$ being the induced metric on $\bE^{(1,\ell)}$, and $m$
is now the induced metric on a section of the worldvolume with constant
$\bE^{(1,\ell)}$ coordinates. Similarly, $\Phi_\ell$ is induced by 
$i_{k_0}\dots i_{k_{\ell}}A$ where $k_0, \dots, k_\ell$ are the 
translational Killing vector fields of $\bE^{(1,\ell)}$. This is the functional
used in \cite{ads} to construct the adS calibration bound. For every
configuration $B$ that minimizes the above energy functional, we find a p-brane
solution with topology $\bE^{(1,\ell)}\times B$.  The considerations of
supersymmetry that we have described above for the $\ell=0$ case  can easily be
extended to the
$\ell>0$ cases. In particular, a generalized calibration satisfies
\bea
\varphi|_\xi&\le& \nu_\ell\, vol_\xi
\nonumber\\
d\varphi&=&{\cal F^\ell}\ ,
\eea
where ${\cal F^\ell}=-di_{k_0}\dots i_{k_{\ell}}A$. 


\section{{\bf M2-BRANES AND GENERALIZED CALIBRATIONS}}

We shall begin by discussing some examples of ordinary calibrations in non-flat
backgrounds. This will allow us to provide an interpretation in terms of
calibrations of some recent work on sigma-model lump solitons on the M2-brane
\cite{bergtow}; specifically we show that they correspond to calibrated
2-surfaces associated to K\"ahler calibrations\footnote{Solutions to the
Cayley calibration equations found in \cite{oct} were characterised
there as `octonionic lumps'.}. We will then turn to the cases
associated with generalized calibrations.

The general setup for lumps on an M2-brane is a probe 
M2-brane in a supergravity
background of the form $\bE^{(1,2)}\times M_8$ with metric
\be
ds^2= ds^2(\bE^{(1,2)}) + ds^2_8\ .
\ee
We embed the worldvolume and fix the worldvolume reparameterizations
such that the $\bE^{(1,2)}$ coordinates are identified with the worldvolume
coordinates. We then consider static worldvolume configurations
$y^a=y^a(\sigma^1,\sigma^2)\}$, where $\{y^a; a=1,\dots 8\}$ 
are coordinates for
$M_8$ and $(\sigma^1,\sigma^2)$ are the worldspace coordinates. The M2-brane
energy is then
\be
E=\int d^2\sigma \sqrt {{\rm det}\, m}\ ,
\ee
where $m$ is the metric induced from $ds^2_8$. We see that, in this case, 
the redshift factor $\nu$ is unity, and $\Phi$ vanishes.

The case considered in \cite{bergtow} was $M_8=\bE^4\times M_4$ with
a KK-monopole metric on $M_4$. This is a circle bundle over $\bE^3$ with
the circle degenerating to a point at the `centres' of the metric, which are
points in $\bE^3$. The subcase considered in \cite{bergtow} was the two-centre
metric, for which there is a privileged direction in $\bE^3$ defined (up to a
sign) by the line between the two centres.  Let $n$ be a unit vector in this
direction. The metric is hyper-K\"ahler so there are three complex structures
$I_r$ ($r=1,2,3$). In particular, the linear combination
\be
J=n^r I_r      
\ee
is a complex structure. The lump configuration of the M2-brane discussed in
\cite{bergtow} is associated with the 2-form 
\be\label{calforma}
\varphi=dx^1\wedge dx^2 +\omega_J\, ,
\ee
where $\omega_J$ is the K\"ahler form of $J$ with respect to the hyper-K\"ahler
metric on $M_4$. This is a K\"ahler form on $\bE^2\times M_4$ and
satisfies the inequality (\ref{bo}). It is therefore a calibration. The
calibration bound is 
$$
E\ge |\varphi|\ ,
$$
which is saturated by the lump configurations. These are desingularized
intersections of the M2-brane probe with an M2-brane wrapped on the finite area
holomorphic 2-cycle of the background. This cycle can be decribed as follows.
Let $\theta$ be the $S^1$ coordinate for the $S^1$ fibre of $M_4$ and choose
the centres in $\bE^3$ of the $M_4$ metric to be at $u=\pm L$ on the u-axis.
Then the 2-cycle is $\{\theta, u; -L\leq u\leq L\}$.

To find the energy of the lump soliton one must subtract the vacuum energy of 
the M2-brane probe. To see how this may be done we note that the energy of the
lump is independent of its size, which is arbitrary. A zero size lump
corresponds to a point intersection of the M2-brane probe with the wrapped
M2-brane. The two M2-branes can then be moved apart, without changing the
energy, so that they `overlap' rather than intersect. We then have two separate
M2-branes, each of which must be a calibrated surface. This is evident for the
probe M2-brane, which is now an infinite planar membrane in $\bE^4$; it is
calibrated by the two-form $dx^1\wedge dx^2$ and its energy is the infinite
vacuum energy. The energy of the lump soliton is therefore 
equal to that of the 
wrapped M2-brane. The energy is finite since the volume of this surface is
finite. The surface itself, described above, is calibrated by the K\"ahler
2-form $\omega_J$. 

Various generalizations of this construction are possible. A multi-centre
metric has various finite area holomorphic 2-cycles around which an M2-brane
may be wrapped. Intersections with another infinite planar M2-brane are again,
when desingularized, lump solitons, although there is now more than one type of
lump. Another generalization
arises from the fact that the 8-metric $ds^2_8$ could be any
eight-dimensional Ricci-flat metric, although it must have reduced holonomy if
the background is to preserve any supersymmetry. The possible reduced holonomy
groups are $SU(2)$, $SU(3)$, $SU(4)$, $G_2$, $Spin(7)$ and $Sp(2)$.
For M2-brane probes in these backgrounds the only relevant calibrations
are of degree two, and the supersymmetric backgrounds admitting
a degree two calibration are those for which the holonomy group is $SU(2)$,
$SU(3)$, $SU(4)$ or $Sp(2)$. All these cases admit
K\"ahler calibrations; the calibrated
surfaces are holomorphic two-dimensional subspaces.

So far we have not needed the full power of {\sl generalized} calibrations.
These arise by placing M2-branes in a D=11 supergravity background
with non-vanishing 4-form field strength. An example is provided by an M2-brane
probe placed parallel to a M2-brane background. The relevant
calibrations that describe the supersymmetric dynamics of the probe are
hermitian of degree two. The  metric $m$ and the electrostatic
potential $\Phi$ of (\ref{enfun}) are induced via the map
$x^i(\sigma)=\sigma^i$, $y^a=y^a(\sigma)$, from the metric 
$ds^2\equiv g_{IJ}dY^I dY^J$ and the 3-form ${\cal F}\equiv -di_kA$, 
respectively, where
\bea
ds^2 &=&H^{-{2\over3}}\left[(dx^1)^2+(dx^2)^2\right] 
+ H^{{1\over3}}\delta_{ab} dy^a dy^b \nonumber\\
{\cal F}&=& dH^{-1}\wedge dx^1\wedge dx^2
\eea
and $H$ is the familiar harmonic function of the multi M2-brane supergravity
solution. The redshift factor is $\nu=H^{-{1\over3}}$. 

To define the Hermitian generalized calibration relevant to this 
case it suffices to give 
the Hermitian form of the complex structure on the 10-space. This
is
\be
\omega_J= H^{-1} dx^1\wedge dx^2 +\omega_{K}
\ee
where $\omega_{K}$ is a (constant) K\"ahler 2-form for
$\bE^8$. Note that ${\cal F}=d\omega_J$, as required for a 
generalized calibration. In addition, $\omega_J$ is the two-form
obtained from the almost complex structure by raising an index with the
rescaled metric
\be
d\tilde s^2= \nu ds^2 = H^{-1}\left[(dx^1)^2+(dx^2)^2\right] +
\delta_{ab} dy^a dy^b\, .
\ee
The almost complex structure is constant and hence integrable. It then
follows from Wirtinger's inequality that $\omega_J$ is a generalized
calibration. The calibrated surfaces are holomophic curves in 
the M2-brane background. Note that the metric $\tilde m$ induced on
worldspace by the rescaled metric $d\tilde s^2$ 
is, in this case, such that $\sqrt{\det \tilde m}=\nu \sqrt{ \det m}$. 

There are many holomorphic curves in the M2-brane background. For
embeddings of the form $x^i=\sigma^i$, $y^a=y^a(\sigma)$, as above,  
they are roughly characterized by the number of `active' transverse
scalars. From the bulk perspective, every pair of active scalars is interpreted
as the worldvolume coordinates of another M2-brane, so the bulk interpretation
of the solution with $2k$ active scalars is that of $k+1$ 
M2-branes intersecting, or overlapping, (not necessarily orthogonally) on a
0-brane soliton, all in the given M2-brane background. This
interpretation suggests that the proportion of the 32 supersymmetries
preserved by the probe M2-brane is $2^{-(k+1)}$, in agreement with 
the proportion derived using the contact set $SU(k+1)/S(U(1)\times U(k))$ 
of the calibration (note that for these examples the presence of the 
background does not affect the count of supersymmetries).

\section{\bf M5-BRANES AND GENERALIZED CALIBRATIONS}

Lumps are also relevant to static M5-brane configurations. For  
this we place a M5-brane in a background of the form $\bE^{(1,5)}\times M_5$
where the $M_5$ metric is Ricci flat and, for preservation of supersymmetry,
has holonomy in $SU(2)$. It follows that $M_5=\bE\times M_4$, with a direct
product metric such that the $M_4$ metric has holonomy in $SU(2)$. The
$SU(2)$ K\"ahler calibration of degree two discussed above is again 
applicable; the interpretation is as M5-branes intersecting on a
3-brane in a KK-monopole background. However, there are now 
other, higher-degree, calibrations to consider. 
One such case
is a degree four K\"ahler calibration, which we now describe. Let $(x^0,x^i)$
($i=1,2,3,4,5$) be cartesian cordinates for $\bE^{(1,5)}$, and let $y^a$
($a=1,\dots, 4$) be coordinates on $M_4$. Consider the manifold $\bE^4\times
M_4$ where $\bE^4$ is the subspace of $\bE^{(1,5)}$ with constant $x^0$ and
$x^5$. This is a K\"ahler manifold with K\"ahler 2-form
\be
\omega=dx^1\wedge dx^2+dx^3\wedge dx^4+\omega_J\ ,
\ee
where $\omega_J$ is defined as for the case of the M2-brane. We fix the
worldvolume diffeomorphisms, and partially specify the embedding of the
M5-brane in spacetime, by choosing $x^0=t$ and identifying $x^i$ with the
worldspace coordinates $\sigma^i$. In addition, we choose the four functions 
$y^a$ to be independent of both the time and one of the worldspace coordinates.
This means that the M5-brane has topology $\bE^{(1,1)}\times B$ where $B$ is a
4-surface parameterised by the other four worldspace coordinates. The four
functions $y^a$ define a minimal  embedding of $B$ in $M_4$ associated to the
calibration 4-form
\be
\varphi={1\over2} \omega\wedge \omega\ .
\ee
If we take $M_4=K_3$ then the M5-brane is wrapped on $K_3$ 
and can be identified
with the heterotic string \cite{harvey}. We have now seen that this 
wrapping can
be described by saying that $B$ is a minimal 4-surface in $K_3$ calibrated by
an $SU(2)$ K\"ahler calibration of degree four. 

Examples of {\sl generalized} calibrations can be found by considering an
M5-brane probe in an M5-brane supergravity background, for which the metric and
seven-form field strength are respectively 
\bea
ds^2&=& H^{-{1\over3}} ds^2(\bE^{(1,5)})+ H^{{2\over3}} ds^2(\bE^5) 
\nonumber\\
F&=&- dH^{{-1}}\wedge dx^0\wedge dx^1\wedge \dots \wedge dx^5\ ,
\eea
where $H$ is a harmonic function on $\bE^5$, which we choose to be
\be
H= 1 + {\mu\over r^3}
\ee
for positive constant $\mu$. We assume here that the
self-dual 3-form field strength on the M5-brane vanishes; otherwise
the theory of generalized calibrations elaborated above would be
inapplicable. The Bianchi identity for this field strength
then implies that the pull-back to the worldvolume of the four form
$F$ must also vanish. This condition is not satisfied for five
active scalars, so we consider only those cases with four or fewer
active scalars\footnote{For a multi-centre M5-brane solution we would
need further restrictions on the number of active transverse scalars
in order to satisfy this condition.}. There are still 
many more such cases than there were for the M2-brane. 
They include `Hermitian',
`Special Almost Symplectic' (SAS) and `exceptional' calibrations, 
generalizing the corresponding calibrations that have been investigated in
the `near-horizon' adS 
backgrounds \cite{ads}. The calibration bounds depends on
the degree of the calibration and the number of active scalars. We are
therefore led to an organisation in which we first specify the number of
active scalars and then consider the various degrees for the calibration form.
In what follows  $\{x^1, \dots, x^{5-\ell}\}$ will denote the worldvolume
coordinates of the M5-brane which are transverse to 
the $\ell$-brane worldvolume
soliton and $\{y^1, \dots, y^n\}$, $n\le 5$, will 
denote the transverse scalars.
As in the M2-brane case, it is convenient to consider a rescaled metric.
Taking this rescaling into account, we have
\bea
d\tilde s^2&=&\nu^{1\over 5-\ell}_\ell 
\left[H^{-{1\over3}} ds^2(\bE^{(5-\ell)})+ H^{{2\over3}}
ds^2(\bE^n)\right]
\nonumber\\
{\cal F}^\ell&=& dH^{-1}\wedge dx^1\wedge \dots \wedge dx^{5-\ell}\ ,
\eea
where $\nu_\ell=H^{-{\ell+1\over 3}}$. The static solutions will be
M5-branes with topology $\bE^{(1,\ell)}\times B$, where $B$ is a calibrated
surface of dimension $p-\ell$. We now consider the various subcases in turn.

\subsection{Hermitian M5-brane Calibrations}

There are four hermitian calibrations
 relevant to the M5-brane which
are best distinguished by the group that 
rotates the complex planes
in their contact set.  In all cases, the 
calibrated submanifold $B$ is a
 holomorphic submanifold
of the M5-brane background.

{\it $SU(2)$ hermitian calibrations}: These are 
degree two calibrations with two
transverse scalars. The relevant four-dimensional 
metric and form field strength are
\bea\label{cfb}
d{\tilde{s}}^2&=&H^{-1}\big[(dx^1)^2+(dx^2)^2\big]+(dy^1)^2
+(dy^2)^2
\nonumber\\
{\cal F}^3&=& dH^{-1}\wedge dx^1 \wedge dx^2\ , 
\eea
respectively.
The  calibration 2-form is
\be\label{cfc}
\varphi=H^{-1} dx^1 \wedge dx^2 +dy^1 \wedge dy^2\ , 
\ee
which is the Hermitian form of a complex structure. The 
complex coordinates are $\{z^1=y^1 +i
y^2, z^2=x^1+ix^2\}$ and the calibrated submanifold is given by the
vanishing locus of a holomorphic function $f(z^1, z^2)$.
The `Killing' spinors obey the conditions
\bea\label{cfe}
\Gamma_0 \Gamma_1 \Gamma_2 \Gamma_3 \Gamma_4 \Gamma_5 \e &=& \e
\nonumber\\
\Gamma_1 \Gamma_2 \e &=& \Gamma_6 \Gamma_7 \e.
\eea 
where the gamma matrices are in an orthonormal basis. We conclude that
the solution preserves  $1 \over 4$ of bulk supersymmetry\footnote{The first of
these conditions can be interpreted as due either to the background or
to the probe, so the examples being considered are those for which the
background does not impose additional constraints.}.
The bulk interpretation of the solutions is that of two M5-branes 
intersecting (at $SU(2)$ angles) on a 3-brane, with one of them
parallel to the background.

{\it $SU(3)$ hermitian calibrations}: There are two cases to
consider. The first is a degree two calibration with four transverse
scalars. The relevant metric and form field strength are, respectively, 
\bea\label{cfg}
d{\tilde{s}}^2&=&H^{-1}\big[(dx^1)^2+(dx^2)^2\big]
\nonumber\\
&+&(dy^1)^2+(dy^2)^2+(dy^3)^2+(dy^4)^2
\nonumber\\
{\cal F}^3&=& dH^{-1}\wedge dx^1 \wedge dx^2\ .
\eea
The calibration two-form is
\be\label{cfh}
\varphi=H^{-1} dx^1 \wedge dx^2 +dy^1 \wedge dy^2+dy^3
\wedge dy^4 \ ,
\ee
which is again the Hermitian form of a complex structure.
The holomorphic coordinates are  $\{z^1=y^1+iy^2, 
z^2=y^3+iy^4, z^3=x^1+ix^2\}$ and the
calibrated surfaces can be described as the 
vanishing locus of two holomorphic
functions $f^1(z^1, z^2, z^3), f^2(z^1, z^2, z^3)$. 
The Killing spinors satisfy the
conditions
\bea\label{cfr}
\Gamma_0 \Gamma_1 \Gamma_2 \Gamma_3 \Gamma_4 \Gamma_5 \e &=& \e
\nonumber\\
\Gamma_1 \Gamma_2 \e &=& \Gamma_6 \Gamma_7 \e
\nonumber\\
\Gamma_1 \Gamma_2 \e &=& \Gamma_8 \Gamma_9 \e\, ,
\eea
from which we deduce that the solution preserves $1 \over 8$ of 
(bulk) supersymmetry. The bulk interpretation
of the configuration is as intersecting M5-branes  
at $SU(3)$ angles for which the corresponding orthogonal intersection is
\bea\label{cff}
&M5& \quad 0,1,2,3,4,5,*,*,*,*
\nonumber\\
&M5& \quad 0,*,*,3,4,5,6,7,*,*
\nonumber\\
&M5& \quad 0,*,*,3,4,5,*,*,8,9.
\eea

The other SU(3) hermitian calibration is a degree four calibration 
with two transverse scalars. The relevant 
metric and form field strength are
\bea\label{cfl}
d{\tilde{s}}^2&=&H^{-{1 \over 2}}
\big[ (dx^1)^2+(dx^2)^2+(dx^3)^2+(dx^4)^2 \big]
\nonumber\\
&+&H^{1 \over 2}\big[
(dy^1)^2+(dy^2)^2 \big].
\nonumber\\
{\cal F}^1&=& dH^{-1}\wedge dx^1 \wedge dx^2 \wedge dx^3 \wedge dx^4\ ,
\eea
respectively.
To describe the calibration form, we introduce the Hermitian form 
\be\label{ethree}
 \omega = H^{- {1 \over 2}} \big[ dx^1 \wedge
dx^2 + dx^3 \wedge dx^4 \big] +H^{1 \over 2}dy^1 \wedge dy^2\ . 
\ee
The corresponding complex structure can be found by raising an index
with the metric defined by $d\tilde s^2$. The calibration form is 
$\varphi={1\over2} \omega\wedge \omega$.
In particular,
\bea\label{efour}
\varphi&=& H^{-1} dx^1 \wedge dx^2 \wedge dx^3 \wedge dx^4
\nonumber\\
&+&dx^1 \wedge dx^2 \wedge dy^1 \wedge dy^2
\nonumber\\
 &+& dx^3 \wedge dx^4
\wedge dy^1 \wedge dy^2\, .
\eea
The holomorphic coordinates are  
$\{z^1=y^1+iy^2, z^2=y^3+iy^4, z^3=x^1+ix^2\}$ and the
calibrated surfaces can be described as 
the vanishing locus of a holomorphic
function
$f(z^1, z^2, z^3)$. The Killing spinors satisfy the
conditions
\bea\label{cfrtwo}
\Gamma_0 \Gamma_1 \Gamma_2 \Gamma_3 \Gamma_4 \Gamma_5 \e &=& \e 
\nonumber\\
\Gamma_1 \Gamma_2 \e &=& \Gamma_6 \Gamma_7 \e
\nonumber\\
\Gamma_3 \Gamma_4 \e &=& \Gamma_6 \Gamma_7 \e\, ,
\eea
so the solution preserves $1 \over 8$ of the bulk supersymmetry. 
The bulk interpretation of the configuration is as intersecting
M5-branes at $SU(3)$ angles for which the corresponding 
orthogonal intersection is 
\bea\label{cfk}
&M5& \quad 0,1,2,3,4,5,*,*
\nonumber\\
&M5&\quad 0,*,*,3,4,5,6,7 
\nonumber\\
&M5&\quad 0,1,2,*,*,5,6,7.
\eea

{\it $SU(4)$ hermitian calibration}: The calibration 
of interest from the M5-brane perspective is
the degree four calibration with  four 
transverse scalars. The relevant metric
and form field strength are, respectively, 
\bea\label{cfq}
 d{\tilde{s}}^2&=&H^{- {1 \over 2}}\big[
(dx^1)^2+(dx^2)^2+(dx^3)^2+(dx^4)^2 \big]
\nonumber\\
&+&H^{1 \over 2}
\big[ (dy^1)^2+(dy^2)^2+(dy^3)^2+(dy^4)^2 \big].
\nonumber\\
{\cal F}^1&=& dH^{-1}\wedge dx^1 \wedge dx^2 \wedge dx^3
\wedge dx^4\ .
\eea
To describe the calibration form we introduce the Hermitian form
\bea\label{esix}
\omega &=& H^{-{1 \over 2}} \big[ dx^1 \wedge
dx^2 + dx^3 \wedge dx^4 \big]
\nonumber\\
 &+&H^{1 \over 2}\big[ dy^1 
\wedge dy^2 +dy^3 \wedge dy^4 \big]\ .
\eea
Then the calibration four-form is 
$\varphi={1\over 2} \omega\wedge \omega$, i.e.
\bea\label{eseven}
 \varphi &=& H^{-1} dx^1 \wedge dx^2 \wedge dx^3
\wedge dx^4
\nonumber \\
&+&\big[ dx^1 \wedge dx^2 \wedge dy^1 \wedge dy^2 
\nonumber\\
&+&dx^1
\wedge dx^2 \wedge dy^3 \wedge dy^4 
\nonumber\\
 &+&dx^3 \wedge dx^4 \wedge
dy^1 \wedge dy^2 
\nonumber\\
&+& dx^3 \wedge dx^4 \wedge dy^3 \wedge dy^4 \big]
\nonumber\\
&+&H dy^1 \wedge dy^2 \wedge dy^3 \wedge dy^4
\eea
The complex coordinates are  $\{z^1=y^1+iy^2, z^2=y^3+iy^4, 
z^3=x^1+ix^2, z^4=x^3+ix^4\}$ and the calibrated manifolds can 
be described as the vanishing locus of two holomorphic 
functions $f^1(z^1, z^2, z^3, z^4), 
f^2(z^1, z^2, z^3, z^4)$. The conditions 
on the Killing spinors are 
\bea\label{cfrthree}
\Gamma_0 \Gamma_1 \Gamma_2 \Gamma_3 \Gamma_4 \Gamma_5 \e &=& \e
\nonumber\\
\Gamma_1 \Gamma_2 \e &=& \Gamma_6 \Gamma_7 \e
\nonumber\\
\Gamma_3 \Gamma_4 \e &=& \Gamma_6 \Gamma_7 \e
\nonumber\\
\Gamma_6 \Gamma_7 \e &=& \Gamma_8
\Gamma_9 \e\, ,
\eea
from which we deduce that the solution preserves $1 \over 16$ of the 
bulk supersymmetry.

In all of these cases it is clear that ${\cal F}^\ell= d \varphi$, 
as required for
consistency. In  addition, in all the above cases an alternative
interpretation of the solutions is as M5-branes with topology
$\bE^{(1,\ell)}\times C$, for either $\ell=1$ or $\ell=3$, where $C$ 
is a $(5-\ell)$-dimensional holomorphic cycle.

\subsection{SAS Calibrations}

The Special Almost Symplectic 
calibrations (SAS) of  $AdS_7\times S^4$, which
is the near horizon geometry of the 
M5-brane, have a natural generalization
in the full M5-brane background. 
The SAS calibrations associated with a M5-brane 
probe placed parallel to a M5-brane supergravity
background have degrees two, 
three, or four (a calibration of degree five would have to involve
the 2-form tensor field and is therefore excluded). The degree two
SAS calibration is identical to the 
degree two Hermitian calibration
explained in the previous section. 
It therefore remains to describe the  SAS
calibrations with degrees three and four. For SAS calibrations 
the number of transverse scalars is the same as
the degree of the calibration. So the degree three SAS 
calibration is a three-dimensional submanifold in
a six-dimensional manifold and the degree four 
calibration is a four-dimensional submanifold
in a eight-dimensional manifold.

For the degree three SAS
calibration, the relevant metric
and form field strength are 
\bea\label{xyz}
d {\tilde{s}}^2 &=&H^{-{2 \over 3}} \big[ (dx^1)^2+(dx^2)^2
+(dx^3)^2 \big]
\nonumber\\
&+&H^{1 \over 3} \big[ (dy^1)^2+(dy^2)^2+(dy^3)^2
\big]
\nonumber\\
{\cal F}^2&=&  dH^{-1}\wedge dx^1\wedge dx^2\wedge dx^3\ ,
\eea
respectively.

To proceed we introduce the orthonormal frame
\bea
e^p&=&H^{-{1\over3}} dx^p\ , \qquad 1\le p\le 3
\nonumber\\
e^{3+p}&=&H^{{1\over6}} dy^p\ , \qquad 1\le p\le 3
\eea
The almost symplectic form is defined as
\be
\omega\equiv \sum^3_{p=1} e^p\wedge e^{3+p}= 
H^{-{1\over6}}\sum^3_{i=1} dx^i\wedge dy^i\ .
\ee
Using the almost complex structure 
associated with this almost symplectic
form, the calibration form is found to be
\bea
\varphi&\equiv&{\rm Re} (e^1+ie^4)\wedge \dots \wedge (e^3+ie^6)
\nonumber\\
&=& H^{-1}dx^1 \wedge dx^2 \wedge dx^3 -dx^1 \wedge dy^2 \wedge
dy^3
\nonumber\\
&-&dy^1 \wedge dx^2 \wedge dy^3 - dy^1 \wedge dy^2 \wedge dx^3
\eea
Repeating the same arguments as in the AdS case in \cite{ads},
the calibrated submanifolds are seen to be 
determined by a single real function
$f(x^1, x^2, x^3)$ such that 
\bea\label{sle}
y^i &=& {\partial\over \partial x^i} f(x^1 ,
x^2 , x^3)
\nonumber\\
H^{-1} \delta^{mn}\partial_{m}\partial_{n}f
&=&{\rm det} (\partial_{i}\partial_{j}f)\ .
\eea

The Killing spinors of such solutions satisfy the conditions
\bea\label{slf}
\Gamma_0 \Gamma_1 \Gamma_2 \Gamma_3 \Gamma_4 \Gamma_5 \e &=& \e
\nonumber\\
\Gamma_0 \Gamma_3 \Gamma_4 \Gamma_5 \Gamma_6 \Gamma_7 \e &=& -\e
\nonumber\\
\Gamma_0 \Gamma_2 \Gamma_4 \Gamma_5 \Gamma_6 \Gamma_8 \e &=& \e\, ,
\eea
and so the configuration preserves  ${1 \over 8}$ of the
bulk supersymmetry. The bulk interpretation of the solution is that of three
intersecting  M5-branes, possibly at $SU(3)$ angles, for which the associated
orthogonal intersection is
\bea\label{sla}
&M5& \quad 0,1,2,3,4,5,*,*,*
\nonumber\\
&{\overline{M5}}&\quad 0,*,*,3,4,5,6,7,*
\nonumber\\
&M5& \quad 0,*,2,*,4,5,6,*,8.
\eea
Alternatively, the solution describes a M5-brane with 
topology $\bE^{(1,2)}\times B$,
where $B$ is a three-dimensional SAS calibrated submanifold.

For the degree four calibration, the 
relevant metric and form field strength
are
\bea\label{cfqtwo}
 d{\tilde{s}}^2&=&H^{- {1 \over 2}}\big[
(dx^1)^2+(dx^2)^2+(dx^3)^2+(dx^4)^2 \big]
\nonumber\\
&+&
H^{1 \over 2}\big[ (dy^1)^2+(dy^2)^2+(dy^3)^2+(dy^4)^2 \big]
\nonumber\\
{\cal F}^1&=&  dH^{-1}\wedge dx^1\wedge \dots \wedge dx^4\ ,
\eea
respectively.
We again introduce the frame
\bea
e^p&=&H^{-{1\over4}} dx^p\ , \qquad 1\le p\le 4
\nonumber\\
e^{4+p}&=&H^{{1\over4}} dy^p\ , \qquad 1\le p\le 4\ ,
\eea
and the  symplectic form
\be
\omega\equiv \sum^4_{p=1} e^p\wedge e^{4+p}= 
\sum^3_{i=1} dx^i\wedge dy^i\ .
\ee
Using again the almost complex structure 
associated with $\omega$, the
calibration form is found to be
\bea
\varphi &=&{\rm Re} (e^1+ie^4)\wedge \dots \wedge (e^4+ie^8)
\nonumber\\
&=&H^{-1}dx^1 \wedge dx^2 \wedge dx^3 \wedge dx^4 
\nonumber\\
&-&dx^1 \wedge
dx^2 \wedge dy^3 \wedge dy^4 
\nonumber\\
&-& dx^1 \wedge dy^2 \wedge dx^3 \wedge
dy^4
\nonumber\\
&-& dx^1 \wedge dy^2 \wedge dy^3 \wedge dx^4
\nonumber\\
 &-& dy^1 \wedge dy^2
\wedge dx^3 \wedge dx^4
\nonumber\\
&-& dy^1 \wedge dx^2 \wedge dy^3 \wedge dx^4
\nonumber\\
&-&dy^1 \wedge dx^2 \wedge dx^3 \wedge dy^4
\nonumber\\ 
&+&H dy^1 \wedge dy^2 \wedge dy^3 \wedge dy^4\ .
\eea
As in the previous case, the solutions are 
determined by a single real function
$f(x^1, x^2, x^3, x^4)$ such that 
\bea\label{slp}
y^i &=&{\partial\over \partial x^i}f(x^1 , x^2 , x^3 , x^4)
\nonumber\\
H^{-1} \delta^{mn}\partial_{m}\partial_{n}f&=&\sum_{m}
{\det}_{m|m}(\partial_{i}\partial_{j}f)
\eea
where $\det_{m|m}$ denotes the determinant of a matrix with the 
m-th row and column omitted.  The Killing spinors 
of these solutions obey the
conditions 
\bea\label{slq}
\Gamma_0 \Gamma_1 \Gamma_2 \Gamma_3 \Gamma_4 \Gamma_5 \e &=& \e
\nonumber\\
\Gamma_1 \Gamma_2 \e &=&- \Gamma_6 \Gamma_7 \e
\nonumber\\
\Gamma_1 \Gamma_3 \e &=&- \Gamma_6 \Gamma_8 \e
\nonumber\\
\Gamma_1 \Gamma_4 \e &=&- \Gamma_6 \Gamma_9 \e
\eea
and so they preserve $1 \over 16$ of the bulk
supersymmetry. The bulk interpretation of the solutions is as 
four intersecting M5-branes, possibly at 
$SU(4)$ angles, for which the
corresponding orthogonal intersection is
\bea\label{sll}
&M5& \quad 0,1,2,3,4,5,*,*,*,*
\nonumber\\
&{\overline{M5}}&  \quad 0,*,*,3,4,5,6,7,*,*
\nonumber\\
&M5&  \quad 0,*,2,*,4,5,6,*,8,*
\nonumber\\
&{\overline{M5}}&  \quad 0,*,2,3,*,5,6,*,*,9.
\eea
Alternatively, the solution describes a M5-brane with topology 
$\bE^{(1,1)}\times B$,  where $B$ is a four-dimensional SAS submanifold.

\subsection{Exceptional Calibrations}

There are three exceptional calibrations to consider as follows:
The Cayley calibration which is a degree four calibration with
four transverse scalars. The relevant eight-dimensional 
metric and form field strength
are
\bea\label{cfqthree}
d{\tilde{s}}^2&=&H^{- {1 \over 2}}\big[
(dx^1)^2+(dx^2)^2+(dx^3)^2+(dx^4)^2 \big]
\nonumber\\
&+&H^{1 \over 2}
\big[ (dy^1)^2+(dy^2)^2+(dy^3)^2+(dy^4)^2 \big]
\nonumber\\
{\cal F}^1&=& dH^{-1} \wedge dx^1\wedge \dots\wedge dx^4\ ,
\eea
respectively.
The calibration four-form can be constructed 
from the metric and a $Spin(7)$ 
invariant self-dual four-form on $\bE^8$, {\it viz}
\bea\label{xyzz}
\varphi &=& H^{-1} dx^1 \wedge dx^2 \wedge dx^3  \wedge dx^4 +\big[ dx^3
\wedge dx^4 \wedge dy^3 \wedge dy^4
\nonumber\\
&+& dx^2 \wedge dx^4 \wedge dy^2 \wedge dy^4 + dx^2
\wedge dx^3 \wedge dy^2 \wedge dy^3 
\nonumber\\
&+& dx^1
\wedge dx^3 \wedge dy^2 \wedge dy^4
+ dx^1 \wedge dx^2 \wedge dy^1 \wedge dy^2 
\nonumber\\
&+& dx^1
\wedge dx^3 \wedge dy^1 \wedge dy^3
+ dx^1
\wedge dx^4 \wedge dy^1 \wedge dy^4
\nonumber\\
&+& dx^2 \wedge dx^4 \wedge dy^1 \wedge dy^3 - dx^1
\wedge dx^2 \wedge dy^3 \wedge dy^4
\nonumber\\
&-& 
dx^1\wedge dx^4 \wedge dy^2 \wedge dy^3
- dx^3 \wedge dx^4 \wedge dy^1 \wedge dy^2 
\nonumber\\
&-& dx^2
\wedge dx^3 \wedge dy^1 \wedge dy^4 \big] \nonumber\\
&& +\ H dy^1 \wedge dy^2 \wedge dy^3 \wedge dy^4\ .
\eea
The equations  which the transverse scalars satisfy
are
\bea\label{egc}
H^{-1}( \partial_{1} Y&-&\partial_{2}Y i-\partial_{3}Y j-\partial_{4}Y
k)=\partial_{2}Y 
\times \partial_{3}Y \times \partial_{4}Y
\nonumber\\
&+&  \partial_{1}Y \times \partial_{3}Y
\times
\partial_{4}Y i
-\partial_{1}Y \times \partial_{2}Y \times \partial_{4}Y j
\nonumber\\
&+&\partial_{1}Y 
\times \partial_{2}Y \times \partial_{3}Y k
\eea\label{egd}
together with the second order auxiliary condition
\bea
{\rm Im}&[&(\partial_1 Y \times \partial_2 Y - 
\partial_3 Y \times \partial_4 Y )i
\nonumber\\
&+&
(\partial_1 Y \times \partial_3 Y + \partial_2 Y \times \partial_4 Y )j
\nonumber\\
&+&(\partial_1 Y \times \partial_4 Y - 
\partial_2 Y \times \partial_3 Y )k]=0\ ,
\eea
where $Y=y^1+iy^2+jy^3+ky^4$, 
$a\times b\times c={1\over2} (a\bar b c-c\bar b a)$, $a\times
b=-{1\over2} (\bar a b-\bar b a)$ and $i,j,k$ 
are the imaginary unit quaternions. The
Killing spinors satisfy the conditions 
\bea\label{ege}
\Gamma_0 \Gamma_1 \Gamma_2 \Gamma_3 \Gamma_4 \Gamma_5 \e &=& \e
\nonumber\\
\Gamma_3 \Gamma_4 \e &=&- \Gamma_8 \Gamma_9 \e
\nonumber\\
\Gamma_2 \Gamma_3 \e  &=&- \Gamma_7 \Gamma_8 \e
\nonumber\\
\Gamma_1 \Gamma_2 \e  &=&- \Gamma_6 \Gamma_7 \e
\nonumber\\
\Gamma_1 \Gamma_4 \e  &=& \Gamma_7 \Gamma_8 \e\, ,
\eea
so the solution preserves $1 \over 32$ of bulk supersymmetry.

The  Associative Calibration is a degree three calibration with four
transverse scalars. The relevant  
seven-dimensional metric and form field strength are 
\bea
d {\tilde{s}}^2 &=& H^{-{2 \over 3}}\big( (dx^1)^2+(dx^2)^2+(dx^3)^2
\big) 
\nonumber\\
&+&H^{{1 \over 3}} \big( (dy^1)^2+(dy^2)^2+(dy^3)^2+(dy^4)^2
\big)
\nonumber\\
{\cal F}^2&=& dH^{-1}\wedge dx^1\wedge dx^2\wedge dx^3\ ,
\eea
respectively
The calibration three-form is constructed from the above metric 
and the structure constants
of  octonions as follows:
\bea\label{aaa}
\varphi &=& H^{-1} dx^1 \wedge dx^2 \wedge dx^3 + 
\big[ dx^1 \wedge dy^1 \wedge dy^2
\nonumber\\
&+&dx^3 \wedge dy^1 \wedge dy^4
+ dx^2 \wedge dy^2 \wedge dy^4
\nonumber\\
&+&dx^2 \wedge dy^1 \wedge dy^3 
-dx^1 \wedge dy^3 \wedge dy^4 
\nonumber\\
&-&dx^3 \wedge dy^2 \wedge
dy^3 \big]\ .
\eea
The equations  which the transverse scalars  satisfy
are
\be\label{egh}
-H^{-1}(\partial_{1} Y i + \partial_{2} Y j + \partial_{3} Y k) = 
\partial_{1} Y \times \partial_{2} Y \times \partial_{3}Y\ ,
\ee
where $Y$ is defined as in the Cayley case.
The conditions on the Killing spinor are 
\bea\label{egi}
\Gamma_0 \Gamma_1 \Gamma_2 \Gamma_3 \Gamma_4 \Gamma_5 \e &=& \e
\nonumber\\
\Gamma_2 \Gamma_3 \e &=&- \Gamma_8 \Gamma_9 \e
\nonumber\\
\Gamma_1 \Gamma_3 \e &=&- \Gamma_7 \Gamma_9 \e
\nonumber\\
\Gamma_1 \Gamma_3 \e &=&- \Gamma_6 \Gamma_8 \e\, ,
\eea
so this configuration preserves $1 \over 16$ of the bulk supersymmetry.

The Coassociative Calibration is a degree four 
calibration with three transverse
scalars. The relevant  seven-dimensional
metric and form field strength are 

\bea
d {\tilde{s}}^2 &=& H^{-{1 \over 2}} \big( (dx^1)^2
+(dx^2)^2+(dx^3)^2+(dx^4)^2 \big) 
\nonumber\\
&+&H^{{1 \over 2}} \big( (dy^1)^2
+(dy^2)^2 +(dy^3)^2 \big)
\nonumber\\
{\cal F}^1&=& dH^{-1}\wedge dx^1\wedge \dots\wedge dx^4\ ,
\eea
respectively.
The calibration four-form is constructed from the above metric and the
dual form of the structure constants of octonions in $\bE^7$ as
\bea
\varphi&=& H^{-1} dx^1 \wedge dx^2 \wedge dx^3 \wedge
dx^4 
\nonumber\\
&+& \big[ dx^3 \wedge dx^4 \wedge dy^2\wedge dy^3
+dx^2 \wedge dx^4 \wedge dy^1
\wedge dy^3
\nonumber\\
&+&dx^2 \wedge dx^3 \wedge dy^1 \wedge dy^2
+dx^1 \wedge dx^3 \wedge
dy^1 \wedge dy^3 
\nonumber\\
&-& dx^1 \wedge dx^2 \wedge dy^2
\wedge dy^3
- dx^1 \wedge dx^4 \wedge dy^1
\wedge dy^2 \big]\ .
\eea
The transverse scalars of this calibration satisfy the equations
\be\label{egl}
-H^{-1}(\partial y^1 i + \partial y^2 j + \partial y^3 k) = 
\partial y^1 \times \partial y^2 \times \partial y^3\ ,
\ee
where $\partial=\partial_1+i \partial_2+j \partial_3+k\partial_4$; 
$\partial_i=\partial/\partial x^i$.  The conditions on the Killing spinors are 
\bea\label{egm}
\Gamma_0 \Gamma_1 \Gamma_2 \Gamma_3 \Gamma_4 \Gamma_5 \e &=& \e
\nonumber\\
\Gamma_3 \Gamma_4 \e &=&- \Gamma_7 \Gamma_8 \e
\nonumber\\
\Gamma_1 \Gamma_2 \e &=& \Gamma_7 \Gamma_8 \e
\nonumber\\
\Gamma_2 \Gamma_4 \e &=&- \Gamma_6 \Gamma_8 \e\, ,
\eea
so the solutions preserve $1 \over 16$ of the bulk
supersymmetry. In all the above three types of calibration,
${\cal F}^\ell= d\varphi$ as required for consistency.

\section{{\bf CONCLUSIONS}}

We have shown that generalized calibrations arise naturally
in the context of supersymmetric configurations of p-brane
actions. In particular, we have derived the generalized calibration 
bound from the p-brane supersymmetry algebra. We have also presented
several examples of such calibrations in M-brane backgrounds.
These examples by no means exhaust the possibilities. An obvious generalization
would be to consider 1/4 supersymmetric, intersecting-brane, backgrounds
determined by two harmonic functions. More generally still, one can
take any supersymmetric intersecting brane configuration in flat
spacetime and replace one of the participating branes by its
corresponding supergravity solution. One then expects to recover the 
full intersecting brane configuration as a generalized calibration of
a probe brane in this background. Our examples are exclusively M-theoretic
but string theory examples can easily be found and many are simply related to
those discussed here by some duality chain. 

The main limitation of calibrations is that the theory is inapplicable in those
cases for which there are `active' worldvolume vectors or tensors, as can
happen for string theory D-branes and the M5-brane. Here again, however, many
such cases are related to calibrations via duality chains so this limitation is
not as severe as might be thought. We should point out that
calibrations were discussed in the D-brane context in \cite{stan}. 
There is also the posibility that a
reformulation of D-branes and the M5-brane in which worldvolume vectors and
tensors are induced from an extended spacetime, as recently proposed
\cite{val}, might allow a direct extension of the theory of calibrations to
these cases too. 

\vskip 1cm
\noindent{\bf Acknowledgments:}   J.G.  thanks EPSRC
 for a studentship.  G.P. is supported by a
University Research Fellowship from the Royal Society.
We are grateful to Jerome Gauntlett for helpful discussions.


\end{document}